\documentclass{article}
\usepackage{arabtex}
\usepackage{utf8}
\setcode{utf8}
\usepackage{arxiv}
\usepackage[utf8]{inputenc} 
\usepackage[T1]{fontenc}    
\usepackage{hyperref}       
\usepackage{url}            
\usepackage{booktabs}       
\usepackage{amsfonts}       
\usepackage{nicefrac}       
\usepackage{microtype}      
\usepackage{lipsum}		
\usepackage{graphicx}
\usepackage{natbib}
\usepackage{doi}

\title{End-to-End Solution for Linked Open Data query-logs Analytics}

\author{ \href{https://orcid.org/0000-0002-3794-844X}{\includegraphics[scale=0.06]{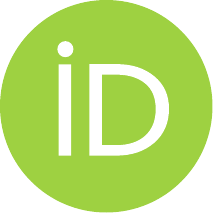}\hspace{1mm}Dihia LANASRI}\\
	ESI\\
	Algiers, Algeria\\	
	\texttt{ad\_lanasri@esi.dz} \\
}

\renewcommand{\shorttitle}{LOD Logs Analytics}

\hypersetup{
pdftitle={End-to-End Solution for Linked Open Data query-logs Analytics},
pdfsubject={NLP, ANALYTICS},
pdfauthor={D.LANASRI},
pdfkeywords={Linked Open Data, query-logs, Layered Architecture, end-to-end solution, Log Analytics},
}

\begin{document}
\maketitle

\begin{abstract}
Important advances in pillar domains are derived from exploiting query-logs which represents users' interest and preferences. Deep understanding of users provides useful knowledge which can influence strongly decision-making. In this work, we want to extract valuable information from Linked Open Data (LOD) query-logs. LOD logs have experienced significant growth due to the large exploitation of LOD datasets. However, exploiting these logs is a difficult task because of their complex structure. Moreover, these logs suffer from many risks related to their Quality and Provenance, impacting their trust. To tackle these issues, we start by clearly defining the ecosystem of LOD query-logs. Then, we provide an end-to-end solution to exploit these logs. At the end, real LOD logs are used  and a set of experiments are conducted to validate the proposed solution. 
\end{abstract}


\section{Introduction}
In big data Era, significant advances in e-commerce, targeted marketing, social shopping, e-tourism, etc. are derived basically from collective intelligence. Such applications mainly exploit data generated by users to extract different valuable information. User content represents data, information, or media content voluntarily provided by people \cite{krumm2008user}, when they interact with web sites, social media, and data sources, etc. This data regroups social data, YouTube videos, blogs and micro-blogs, query-logs, etc. 

Analysis of this data provides useful information helping to understand user behavior, user opinions, topics of interest, etc. It helps to detect hidden patterns and to construct users’ profiles, in order to propose user-centric solutions like: recommendation systems, content personalization, cache improvement, etc. for successful user experience.

Query-logs are important user content which has been considered in literature to tackle many issues. Query-logs record users' manipulations over a data source, traced as textual lines in log files. 

To extract meaningful information from query-logs, they should be cleansed. In their raw format, these logs suffer from serious quality and trust issues \cite{lanasri2020trust}. Consequentially, different curation solutions have been proposed to clean query-logs and store them in adequate data stores for future exploitation.

In this work, we will be interested in Linked Open Data (LOD) query-logs. LOD have experienced significant growth in the industrial and academic worlds because of their openness to public on the web. The adoption of these LOD by large manufacturers is due to the services they provide.

The wide exploitation of LOD datasets generates a large amount of SPARQL query-logs. These logs represent interest of LOD consumers. They have interested the research community for many years and for different purposes like: Recommendation \cite{chen2014sparql}, statistical analysis \cite{bonifati2020analytical}, source selection \cite{tian2011enhancing}, analytical purposes for decision-making; an analysis perspective \cite{khouri2019loglinc}.

As stated previously, in order to ensure an efficient exploitation of any source query-logs, an efficient curation process is required. Moreover, we believe that these logs should be curated under a Trust perspective \cite{lanasri2020trust}. Trust is a complex concept which is linked to risk, quality and provenance. The openness and collaborative characteristics of these logs considerably decreases their trust, since they suffer from many risks related to their Quality  \cite{ceolin2015linking} and Provenance \cite{suriarachchi2016crossing}.

Trust have been studied mainly in LOD/KB datasets, because their openness and rapid growth have raised several issues linked to their ownership and quality. These facts have raised several uncertain LOD/KB \cite{DjebriTG19} containing information that are not highly reliable. Trust was considered when conceptualizing these LOD/KB \cite{djebri2019linking}. However, it is not clearly considred for LOD query-logs. 

Analyzing the literature about LOD query-logs indicates that many efforts are dedicated either to ensure a \textit{curation} process or focusing on the \textit{usage} of these query-logs. However these efforts are made in an isolated way, which shows a serious lack of an end-to-end solution for LOD query-logs analytics. 

This motivates us for defining, an end-to-end solution based on a layered architecture for preparing and analyzing LOD query-logs, where trust is considered as a main concern. The layered architecture composed of four layers (Raw query-logs, Preparation and Curation, Storage  and analytical Layers). A set of experiments are conducted to validate our architecture where real LOD query-logs are considered.

This paper is organized as follows: Section 2 describes our related work. Section 3 defines the LOD query-logs ecosystem. Section 4 presents in detail our solution. Section 4 reviews our experiments. Section 5 concludes the paper.

\section{Related Work}
LOD query-logs are widely considered in literature to extract knowledge. In this related work, we will provide an overview of some LOD query-logs usage. 

\begin{enumerate}
\itemsep=0pt
\item Intrusion detection: Detecting robotic and organic queries \cite{malyshev2018getting}
\item LOD query-logs analysis and information extraction for understanding their graphic representation and discovering their inherent characteristics \cite{bonifati2020analytical} using some visualisation tools and graphical user interfaces like DARQL \cite{bonifati2018darql}, and SEMLEX \cite{mazumdar2011semlex}. This later proposes a semantic analysis of the contents of query-logs.
\item Query Optimization: Improving source selection using a data mining model, which tries to estimate the minimum number of sources from query-logs and keep only those responding to the given request \cite{tian2011enhancing}.  
\item Recommender systems for queries suggestion \cite{chen2014sparql} based on collaborative filtering knowledge to support the user while building her query.
\item Query reformulation using aggregated graph patterns ranking techniques \cite{rafes2018designing}.
\item Personalization and improvement of the cache data management of SPARQL endpoints \cite{akhtar2020cache}. 
\item Business Intelligence: Exploring multidimensional (MD) patterns from open LOD query-logs for data analytics that we proposed in \cite{khouri2019loglinc} using an interactive dedicated tool \cite{lanasri2019crumbs4cube} for generating a Data warehouse.
\end{enumerate}

The review of the different works about LOD logs shows that minimal preparation and curation are performed to structure the logs \cite{mazumdar2011semlex}. These operations can be classified into: 
\\
(1) Cleaning operations like deduplicate queries \cite{ell2011deriving}, delete wrong queries based on QoS metadata \cite{mazumdar2011semlex,ell2011deriving}, extract select queries \cite{mazumdar2011semlex,ell2011deriving}, RDF Triples extraction \cite{mazumdar2011semlex}. \\
(2) Transformation operations like parsing LOD query-logs using jena \cite{mazumdar2011semlex}, prefix identification for incomplete queries \cite{ell2011deriving}, RDF triple features extraction \cite{elbedweihy2011identifying} and semantic and syntactic SPARQL errors correction \cite{almendros2017detecting}. \\
(3) Merging/integration: as opposite to query-logs of data repositories and the web, merging and integration of LOD query-logs are not treated in literature.

However, even if trust was widely considered in LOD/KB datasets \cite{}, their query-logs may contain many issues that can affect their trustworthiness. Many works tried to tackle some trust related problems like simple preprocessing of SPARQL queries, semantic and syntactic analysis of SPARQL queries \cite{almendros2017detecting} to clean them and enhance their quality, detect bots queries and verify their provenance, etc. which help to improve indirectly log trust but it was not studied under the Trust vision. Moreover, this presentation identifies the lack of works dealing simultaneously with trust and providing an end to end solution for LOD logs analytics exploitation. All These facts motivate our proposal.

\section {LOD Logs Structure and Ecosystem}
Exploiting LOD query-logs is a tedious task linked mainly to their complex ecosystem. To understand this complexity, it is prominent to define this ecosystem and explicit its components. 

A LOD data source is composed of millions of RDF triples <Subject Predicate Object> and associated to an ontology defining its concepts for semantic reasoning. A LOD is fed by many data sources and it belongs to one or many domains covering many topics like Scholarly data LOD which is specific to academic research domain and it contains workshops, authors, papers, etc. topics. This LOD is created by a producer which may be a well-known organism like Facebook or unknown one. A LOD is stored on a given Triple store like Virtuoso which is accessed or consumed by many consumers using different services like Question-Answering systems, chatbots and SPARQL endpoints.   

The large exploitation of LOD datasets by consumers generates query-logs which are the second component of the LOD ecosystem. These logs are composed of many raw query lines, each one represents i) a SPARQL query belonging to a given type (select, describe, construct..) and composed of a Basic Graph Pattern (BGP) which is a set of RDF triple patterns < S P O> = <Subject Predicate Object> and ii) metadata like IP adress, userID, execution datetime. These LOD logs are provided by different providers like USEWOD\footnote{\url{http://usewod.org/}} and LSQ\footnote{\url{http://lsq.aksw.org/}} initiatives.   

The LOD dataset and logs consumers may have different expertise levels, they may be beginners, intermediates or experts in using SPARQL query language and they may have different behaviors (human behavior, or bot malicious behavior) which may affect strongly the trust of LOD ecosystem.

\section{Our proposed End-to-End Solution}
LOD query-logs present serious trust issues making their direct use, without any preprocessing, risky \cite{lanasri2020trust}. According to our previous work \cite{lanasri2020trust}, they suffer from many risks related to their quality and provenance. 

Moreover, these logs are usually generated by unknown and less credible users with many intentions, who may provide inaccurate queries. They may also have many profiles: bots or real users (human) with good or malicious purposes which affect logs provenance. These users may have different expertise levels (level of mastering SPARQL language or RDF Dataset). They may be beginners, experts or intermediates when writing queries which may affect strongly the quality of queries. Management of trust in LOD logs is required in order to exploit  these logs for analytics usage purposes.

\begin{figure}
  \centering
  \includegraphics[width=.95\textwidth]{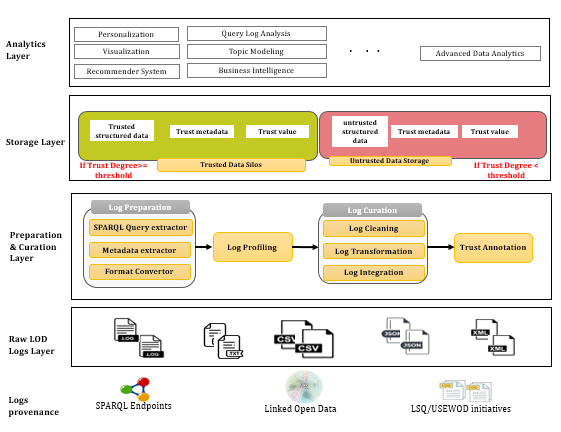}
  \caption{End-to-End Architecture for LOD logs analytics}
  \label{architecture}
\end{figure}

To achieve these goals, we define a trust based layered architecture for LOD query-logs analytics as illustrated in Figure \ref{architecture}. This architecture is composed of four layers (Raw LOD logs Layer, Preparation and Curation Layer, Storage Layer and Analytics Layer) where Trust is projected on each layer. Layered architectures are very familiar in context of data-driven solutions like the multi-layer architecture widely used in Data lake solutions; in software model-driven solution, semantic web layered architecture, big data driven solutions...In what follows, we will detail each layer.

\subsection{Layer 0: Raw LOD Logs Layer} 
The raw  LOD logs should be saved in a physical storage (Data Lake) for future use. These log files may be provided in different formats like JSON, XML and CSV.
According to user requirements and the usage case purpose, these logs are prepared, curated before using them.

\subsection{Layer 1: Preparation and Curation Layer}
To extract just what is needed for the analytics purposes (\textit{bring just what you need}), this layer allows extracting the needed data, fields and the metadata from the appropriate log files from layer 0.

\subsubsection{\textbf{Log Preparation}}
This layer proposes three preparation operators: \\
(i) SPARQL Query extractor which returns the \underline{select} and \underline{construct} SPARQL queries. These two types of queries are kept because they represent the analytical purposes of users; \\
(ii) Metadata extractor returns the different metadata: IP address, Execution datetime, User ID...\\
(iii) Query parser converts the extracted data to a human readable format using UTF-8 parser.

Since these logs are generated by users with different levels of SPARQL language mastery, different profiles and intentions, these logs suffer from several quality problems (ex. false syntax, missing values, etc.) and security problems (like bot generated queries, vulnerable queries, etc). Consequentially, before using them, it is necessary to profile then clean them. 

\subsubsection{\textbf{Log profiling}}
The process of curation starts by profiling logs, it consists of deeply analyzing logs and return statistics about them (rate of duplications, statistics about their quality, their trust, etc.). Log profiling allows: 
\\
\textbf{Log provenance analysis:}
In order to encourage analysts to trust open LOD logs, their provenance should be deeply analyzed. Two main points should be considered Provenance of logs and Provenance of queries.
Users writing queries may have good intentions like learning SPARQL or exploring $LOD$ sources. They may be experts or beginners, but they could sometimes have bad intentions. To identify risky ones, we used operators detailed in \cite{lanasri2020trust}: 

\textit{1- Trusted or vulnerable provider:} detects malicious IP adresses which defines the query provider. \\
\textit{2- Provenance profiling:} identifies the Expertise level of users writing the queries and the Provenance organisms aiming to classify the organism provider into company, academic institution or private user. \\
\textit{3- Behavior analysis:} because IP address is not sufficient to discover user's profile, this technique is proposed to identify organic and bot malicious queries. 

\textbf{Log quality analysis:}
When trust is related to quality \cite{ceolin2015linking} this point needs more attention. Three types of analysis are distinguished: 

\textit{1- Single query analysis:} queries are analyzed one by one to detect: (a) Syntactic errors; (b) Semantic errors (c) Query type: analytic or standard queries and (d) Query complexity: to analyze its depth and shape. Complex queries indicate generally an expert profile behind, which is more trusted.

\textit{2- Analysis of queries interactions:} after analyzing queries as singletons, the interactions between them should be considered. The interactions between queries help to understand their behavior. This consists to detect (a) Duplicate queries, (b) Schema overlap and (c) Topic overlap: to get an idea about the main domains of interest in a LOD log.

\textit{3- Analysis of logs interactions:} the interaction between LOD logs of different sources is considered to identify : (a) Semantic overlap and (b) Sources overlap.

\subsubsection{\textbf{Log Curation}}
Two types of trust based curation operations detailed previously in \cite{lanasri2020trust} are provided, when orchestrated, they form a curation pipeline to ensure data quality and trust. 

\textit{1- Cleaning operators:} these operators eliminate irrelevant data: (a) Robot Query cleaner is used to discard all bot queries not generated by humans; (b) Business/Academic query extractor helps, using WHOSIP\footnote{\url{https://who.is/}}, to select business queries generated by professionals or academic ones generated from academic institutions; (c) Vulnerable query eliminator detects all vulnerable queries that are generated by IPs appearing in a database of blacklisted IPs\footnote{\url{https://github.com/whois-server-list/whois-api-java}}; (d) Deduplicator is used to keep unique queries; (e) Complexity filter detects shapes and depths of queries using \cite{bonifati2018darql} solution, complex queries indicate generally an expert profile behind. (f) Analytic/standard query selector allows selecting standard queries or analytic ones containing aggregate functions that reflect an analysis aim; (g) Topic clustering and (h) Schema ranking are used to detect the topic of a given query based on a created reference base, then deduplicates queries based on the similarity of their triples <S P O>, the results of complexity analysis and the behavior of user to enhance her query, (i) Expertise filter identifies expert from beginner or intermediate profiles.  

\textit{2- Transformation Operators:} They allow correcting some errors using (i) syntactic and (ii) semantic correctors which help to correct wrong queries based on a REGEX and the algorithm proposed by \cite{almendros2017detecting}. 

\subsubsection{\textbf{Trust Annotation}}
After cleaning, the queries will be annotated with a trust degree using the given formula. Then, the data analyst can decide at the end to keep just queries with \\$trustDegree$ > $y$. Where $y$ is a threshold value defined by the data analyst. For instance, keep queries with $trustDegree > 0,75$ after cleaning.

Trust Degree is the probability of trust of a query. $TrustDegree \in [0 , 1]$. We consider the Trust degree formula as
\begin{equation}
 TrustDegree = \frac{1}{NB\_parameters} \sum_{i=1}^{NB\_parameters} (f(x_{ij}))
\end{equation}

where:
$NB\_parameters$  is the number of used operators to analyze queries during log curation. The total number of proposed operators.

$x\_{ij}$ represent the categorical value (annotation) associated with a query when parsed by a given operator analyzer (i). $x\_{i} \in \{x\_{i1}, x\_{i2}...x\_{in}\}$ 

For example, Behavior analysis is the operator analyzer number $i=1$ and $x_{ij}$ associated with this operator are $x_{1j} \in \{bot, organic\}$. \\
If the query is bot $f(x_{ij})=0$ else $f(x_{ij})=1$

The result of this step is two sets of queries: trusted queries (most credible) and untrusted queries. The decision of trusted or untrusted query is based on a the trust threshold defined by the data analyst. 

\subsection{Layer 2: Storage Layer}
Once preprocessed and given their importance, these curated logs must be stored for the various usage cases. The processed LOD logs in general are stored on physical media either on files, data bases or data lake, or in the cloud storage in these recent years. These storage contains the cleansed queries and annotated with a trust degree up to a given threshold. It can also contain untrusted queries for future use like security issues detection.

\subsection{Layer 3: Analytics Layer} 
Once LOD logs are cleansed and stored in the appropriate data storage, this layer allows consuming and exploiting them for a defined purpose or analytics usage case. Many analytics usage cases can be defined in this layer like: Business Intelligence solutions (augmented DW, log DW, Multidimensional exploratory analysis...) or for advanced data analytics solutions based on machine learning and data science...This last layer presents also visualization tools for end-users to explore and analyze the resulted data. 

\subsection{Experiments and Results}
Our experiments are motivated by real case application in academic context. We propose to exploit LOD logs in order to generate Data warehouse of logs for decision makers. Our experiments will be projected on the proposed architecture layer by layer. The series of experiments are executed on a machine OS Windows 10x64 with 16 GB RAM and Intel® core™ i7-3632QM, @ 2.20 GHz CPU using our developed solution in Java and Scala.

\textbf{Layer 0: Raw Logs: }
We selected two different logs for our experiments: (i) Scholarly data log (.log file) provided by LSQ\footnote{\url{https://drive.google.com/file/u/1/d/0B1tUDhWNTjO-T3BweE9YeE1rUGM/view?usp=sharing}}. Scholarly data\footnote{\url{http://www.scholarlydata.org/}} provides data about conferences, workshops and scientific publications. This log contains 5.499.797 raw queries (a set of SPARQL queries and GET/SET queries). (ii) DBpedia logs (.log files) contains many topics about music, films, geography, etc. against DBpedia  3.5.1 LOD\footnote{\url{https://wiki.DBpedia.org/services-resources/datasets/data-set-35/data-set-351}}. In our context, we selected just academic contextualized queries. This log is provided by LSQ\footnote{\url{https://drive.google.com/file/d/0Bw1get4GUTJrWFluNVVhVjAzcjg/view}}. It contains 3.193.672 raw queries where 43.284 are academic queries. Which are stored in a local server (physical storage).

\textbf{Layer 1: Preparation and Curation}
To answer our requirements, the SPARQL queries and the needed metadata (ip address(scholarly data log) and user ID (DBpedia log), execution datetime,  and http-response) are extracted and parsed for future use in next layers.

Then, we execute our trust-aware curation pipeline composed of many transformation, cleaning and integration operators on these logs. 
The process starts by executing a log profiling. The profiling helps to deeply analyze LOD logs and understand their structure, type and get some statistics about them.

After that, we execute the Trust-aware curation pipeline which orchestrates the different operators proposed above in this order: Robot query cleaner -> Business/Academic query extractor -> vulnerable query eliminator -> deduplicator -> syntactic \& semantic correctors -> Topic clustering -> schema ranking -> complexity filter -> analytic/standard query selector.

In our experiments, the different queries are annotated after curation with a Trust Degree using the Trust formula proposed in subsection 4.2.4. 

\begin{table}[!]  
    \centering
    \caption{Scholarly Data and DBpedia Logs curation and MD Patterns}
    \scalebox{0.7}{
    
    \begin{tabular}{@{} l|l|l||c|c @{}}
    \toprule  
        Layer             & Step                        & Types                   &  Scholarly Data                & DBpedia (Academic)    \\ \hline
        Raw Storage       & Raw logs                    & Raw logs                & 5.499.797                      & 3.193.672             \\ \hline
        Preparation       & Academic queries            & Total                   & 5.499.797                      & 43.284                \\ 
                          & SPARQL query type           & Select                  & 141.061                        & 41.138                 \\ 
                          &                             & Construct               & 880                            & 2.146                     \\\hline
                          & Metadata                    & Common                  & Execution DateTime, http code                            \\
                          &                             & Specific                & IP Address, response size      & user session             \\\hline
        Curation          & Log profiling Report                                                                                        \\\hline 
                          & Behavior analysis           & Robot                   & 3.405                          & 4.579  \\ 
                          &                             & Organic                 & 138.536                        & 38.704  \\ \hline
                          & Provenance organism         & Business                & 29.340                         & 0  \\ 
                          &                             & Academic                & 111.088                        & 0  \\ 
                          &                             & Unknown                 & 1513                           & 43.284 \\ \hline
                          & Trusted/Vulnerable Provider & Unknown                 & 0                              & 43.284  \\ 
                          &                             & Vulnerable              & 71                             & 0 \\ 
                          &                             & Safe                    & 141.870                        & 0  \\ \hline
                          & Duplicate queries           & Duplication             & 1.572                          & 6.782  \\
                          &                             & Unique                  & 140.369                        & 36.502  \\ \hline
                          & Syntactic Errors            & Wrong                   & 49.241                         & 5.478  \\ 
                          &                             & Correct                 & 92.700                         & 37.806  \\ \hline
                          & Semantic Errors             & Wrong                   & 12.450                         & 2.001 \\ 
                          &                             & Correct                 & 129.491                        & 41.283  \\ \hline
                          & Topic Overlap               & None                    & 46.692                         & 18  \\ 
                          &                             & Academic Event          & 39.833                         & 2.664  \\ 
                          &                             & Agent                   & 15.925                         & 0  \\ 
                          &                             & Call For                & 77                             & 0  \\ 
                          &                             & Document                & 2.170                          & 1.208 \\ 
                          &                             & Institute               & 16.803                         & 35.676  \\ 
                          &                             & Non-Academic Event      & 413                            & 0  \\ 
                          &                             & Publication             & 25                             & 1.297  \\ 
                          &                             & Role                    & 2.877                          & 364  \\ 
                          &                             & Site                    & 355                            & 13  \\ 
                          &                             & Track                   & 1.638                          & 630 \\
                          &                             & Topic                   & 15.133                         & 1.414  \\ \hline
                          & Schema Overlap              & Informative             & 26.850                         & 16.464  \\ 
                          &                             & Non Informative         & 115.091                        & 26.820 \\ \hline
                          & Query Complexity            & Simple                  & 51.201                         & 19.792 \\ 
                          &                             & Chain                   & 2.935                          & 8  \\ 
                          &                             & Star                    & 87.303                         & 23.438 \\ 
                          &                             & Tree                    & 120                            & 11  \\ 
                          &                             & Flower                  & 152                            & 16  \\ 
                          &                             & Bouquet                 & 154                            & 8  \\ 
                          &                             & Forrest                 & 76                             & 11  \\ \hline
                          & Expertise Level             & Beginner                & 135.307                        & 43.072 2 \\ 
                          &                             & Intermediate            & 4.633                          & 201 \\ 
                          &                             & Expert                  & 2.001                          & 11  \\ \hline
                          & Query Type                  & Analytic                & 5.673                          & 1.381  \\ 
                          &                             & Standard                & 136.268                        & 41.903  \\ \hline
        Storage           & Trusted Data                & Trusted Queries         & 114.972                        & 28.134    \\
                          & Untrusted Data              & Untrusted Queries       & 2.847                          & 15.150    \\ \hline
        Usage             & MD Patterns                 & Facts                   & 10                             & 190    \\
                          &                             & Dimensions              & 15                             & 243    \\ 
                          &                             & Dimension Attributes    & 37                             & 293    \\ 
                          &                             & Fact Attributes         & 0                              & 0    \\ 
                          &                             & Measures                & 10                             & 190    \\  \hline 
        
         \bottomrule
    \end{tabular}
    }
    
    \label{TestProfiling}
\end{table}

In the table\ref{TestProfiling}, we have all the details and statistics about each log. 
We noticed that the status of vulnerability in DBpedia queries is Unknown because these queries don't have an IP address; they contain ID which is not informative. Consequently, we cannot classify them to business or academic. For this, we consider them as Business and Academic at the same time. 

For each log, we execute a pipeline composed of our operators and respecting these parameters. We consider just organic, academic, safe, Unique, syntactically and semantically correct and most informative queries (see table \ref{TestProfiling} for more detailed parameters values). In our case, we want to keep all queries with trust degree more than 0,75.

for example, $Q_1$ is annotated like this (bot , academic , safe , Unique , synt-correct , sem-corrected , document , not Informative, star, intermediate, standard).
Based on our defined preferences and according to the annotations above:

Trust degree ($Q_1$)= 1/11 * (0 + 1 + 1 + 1 + 1 + 1 + 1 + 0 + 1 + 1 + 1) = 9/ 11 = 0,81 > 0,75 so $Q_1$ is accepted  

$Q_2$ is annotated like this: (bot , academic , vulnerable , duplicated , synt-wrong , sem-wrong , role , not Informative, star, beginner, standard).
Based on our defined preferences and according to the annotations above:

Trust degree ($Q_2$)=1/11 *(0 + 1 + 0 + 0 + 0 + 0 + 1 + 0 + 1 + 1 + 1) = 5/ 11 = 0,45 < 0,75 so $Q_2$ is rejected

After curating then annotating all queries (à posteriri annotation), we found that  81\,\% of Scholarly data and 65\,\% of DBpedia queries have a trust degree more than 0,75 while the rest has a trust degree less than 0,75.

The AVG (trustDegree) is 0,89 in Scholarly data log and 0,78 in DBpedia log .

Min(trustDegree) is 0,63 in Scholarly data log and 0,1 DBpedia log.

Max(trustDegree) is 1 in two logs. 

\textbf{Layer 2: Storage}
Once these LOD logs are curated and we deleted untrusted queries, we stored the trusted ones in SQLite database. This database is used to store final queries and the intermediate resulted queries after each operations. Jena TDB triple store is also used to store LOD ontologies.

\textbf{Layer 3: Analytics usage:} 
We used a previous work (Scenario 3: Enrichment of the MD Schemes) \cite{khouri2019loglinc} to explore multidimensional (MD) patterns before generating the log DW. For each log, we run the approach LogLinc \cite{khouri2019loglinc} to extract the multidimensional star schemes. After that, we used semantic similarity (based on Wordnet KB and using WS4J\footnote{\url{https://github.com/Sciss/ws4j}} library) to group similar MD patterns (Facts, Dimensions, etc).  At the end, with the help of BI specialist, we manually checked the matching of these schemes and generate a DW.

This DW is used to achieve the defined goals detailed in beginning of this section. This DW can be used by decision makers to formulate their decision queries to get needed information and make decisions.

We selected the main MD patterns which are: 195 number of fact classes,  247 number of dimension classes, 0 number of fact attributes of the fact classes, 195 number of measures of the fact classes, 304 number of dimension attributes of the dimension classes (Table \ref{TestProfiling}) 

Our proposed architecture participates strongly to exploit LOD logs and insure a good level of trust. Based on the treated and trusted logs, we could generate trusted data warehouses containing relevant data that interest the community, because logs represent the real interest of end users. This can encourage decision makers to exploit these logs via a DW to analyze data and make decisions linked to the presented problematic.

\section{Conclusion}
We proposed an end-to-and solution where a layered architecture is definied and trust dimension is considered to exploit LOD query-logs and extract their valuable knowledge. These logs when well prepared, cleansed and controlled to avoid using risky logs are considered as an important assets for companies and rich source of information which can be exploited for many analytics usage cases. 

The experiments conducted on DBpedia and Scholarly data have supported our proposal and show that our solution is effective to clean these, to ensure their trust and use them in a trustfully way for different analytics usage cases, like generating DW model.

As perspective, we work on developing a Trust based tool supporting our proposal. We want also to analyse the effectiveness of LOD logs in wide data analytics context of companies.

\section{Acknowledgments}
We extend our gratitude to Professor BELLATRECHE Ladjel and Dr. KHOURI Selma for their invaluable guidance, insightful ideas, and contributions.

\bibliographystyle{unsrtnat}
\bibliography{references}

\end{document}